\def\rfr#1{eq.(\ref{#1})}
\def\rfrs#1#2{eqs.(\ref{#1})-(\ref{#2})}

\def\eqi{\begin{equation}}
\def\eqf{\end{equation}}
\def\eqia{\begin{eqnarray}}
\def\eqfa{\end{eqnarray}}

\def\rp#1#2{{#1\over#2}}

\def\lb#1{\label{#1}}

\def\bm#1{{\mbox{\boldmath$#1$\unboldmath}}}

\def\to{$T_{\rm obs}$}

\documentclass[onecolumn]{aa}
\begin{document}

\title{On the possibility of measuring the solar oblateness and
some relativistic effects from planetary ranging}

\author{L. Iorio\inst{1}}

\offprints{L. Iorio}

\institute{Dipartimento di Fisica dell'Universit${\rm \grave{a}}$
di Bari, Via Amendola 173, 70126, Bari, Italy\\
\email{Lorenzo.Iorio@libero.it} }

\date{Received , 2004; accepted , 2004}

\abstract{In this paper we first calculate the post--Newtonian
gravitoelectric secular rate of the mean anomaly of a test
particle freely orbiting a spherically symmetric central mass.
Then, we propose a novel approach to suitably combine the
presently available planetary ranging data to Mercury, Venus and
Mars in order to determine, simultaneously and independently of
each other, the Sun's quadrupole mass moment $J_{2\odot}$ and the
secular advances of the perihelion and the mean anomaly. This
would also allow to obtain the PPN parameters $\gamma$ and $\beta$
independently. We propose to analyze the time series of three
linear combinations of the observational residuals of the rates of
the nodes $\dot\Omega$, the longitudes of perihelia $\dot\pi$ and
mean anomalies $\dot\mathcal{M}$ of Mercury, Venus and Mars
suitably built up in order to absorb the secular precessions
induced by the solar oblateness and the post-Newtonian
gravitoelectric forces. The values of the three investigated
parameters can be obtained by fitting the expected linear trends
with straight lines, determining their slopes in arcseconds per
century and suitably normalizing them. According to the
present-day EPM2000 and DE405 ephemerides accuracy, the obtainable
precision would be of the order of 10$^{-4}$--10$^{-5}$ for the
PPN parameters and, more interestingly, of $10^{-9}$ for
$J_{2\odot}$. It must be pointed out that the future BepiColombo
mission should improve the knowledge of the Mercury's orbit
perhaps by one order of magnitude.


}
\titlerunning {On the Sun's oblateness and the post-Newtonian gravitoelectric field}

\maketitle

\section{Introduction}
In this paper we will deal with certain Newtonian and
post-Newtonian secular effects affecting the nodes $\Omega$, the
longitudes of the perihelia $\pi$ and the mean anomalies
$\mathcal{M}$ of the Solar System's planets.

Historically, one of the first classical tests of the Einstein
General Theory of Relativity (GTR) was the successful explanation
of the anomalous secular perihelion advance of Mercury in the
gravitational field of Sun (Einstein 1915). As we will see, this
feature of the planetary motion is strictly connected with the
problem of the quadrupole mass moment $J_{2\odot}$ of Sun. In
(Pireaux and Rozelot 2003) a theoretical range $J_{2\odot}=(2\pm
0.4)\times 10^{-7}$ is admitted for the Sun's oblateness. For this
topic and the interplay between both effects see the recent review
(Pireaux \& Rozelot 2003) and the references therein and
(Ciufolini \& Wheeler 1995). Basically, the point is the
following. In regard to the secular orbital motions, to which we
are interested here, the Sun's mass quadrupole moment induces
classical effects which have, qualitatively, the same temporal
signature of the relativistic ones. This means that they could
corrupt the recovery of the genuine post-Newtonian features of
motion because they could not be removed from the time series
without removing the post-Newtonian signal of interest as well.
All depends on the precision with which $J_{2\odot}$ is known:
should the mismodelling in it induce classical residual
precessions larger than the relativistic ones, there would  be no
hope to get a reliable test of relativistic gravity. As we will
see later, the same problems could come from the $N-$body secular
precessions. In this paper we propose to disentangle such effects
by measuring them in an independent way. More precisely, we
propose to extend a certain approach used in Earth artificial
satellite motion analysis to the interplanetary arena in order to
single out just certain post--Newtonian (and Newtonian) orbital
motion features independently of the
Parameterized--Post--Newtonian (PPN) framework (Will 1993) which
is usually employed in testing competing metric theories of
gravity. Instead, the current approach consists of testing the
post--Newtonian equations of motion as a whole in terms of the PPN
parameters which are determined from multi--parameters fits
together with other astrodynamical quantities.

In Table \ref{param} the orbital parameters of the inner planets
of the Solar System are reported.
\begin{table}
  \caption[]{Orbital parameters of Mercury, Venus and Mars
(\texttt{http://nssdc.gsfc.nasa.gov/planetary/factsheet/}). For
the Astronomical Unit (A.U.) we use the value 1 A.U.=149597870691
m of the DE405 (Standish 1998) and EPM2000 ephemerides (Pitjeva
2001a; 2001b). The angle $\epsilon$ refers to the inclination to
the ecliptic.}
     \label{param}
$$
\begin{array}{p{0.5\linewidth}lll}
            \hline
            \noalign{\smallskip}
Planet  & a\ (A.U.) & \epsilon\ (^\circ) & e \\
\noalign{\smallskip}
            \hline
            \noalign{\smallskip}
Mercury & 0.38709893 &  7.00487 & 0.20563069 \\
Venus & 0.72333199 & 3.39471 & 0.00677323 \\
Mars & 1.52366231 & 1.85061 & 0.09341233 \\
\noalign{\smallskip}
            \hline
         \end{array}
     $$
\end{table}
\section{The post-Newtonian effects}
\subsection{The gravitoelectric effects}
In the framework of the standard PPN formalism the post-Newtonian
gravitoelectric acceleration induced by the Schwarzschild--like
part of the spacetime metric and experienced by a test body freely
falling around a static, spherically symmetric central mass $M$ is
(Soffel 1989) \eqi \bm a_{\rm GE}=\rp{GM}{c^2
r^3}\left\{\left[2(\gamma+\beta)\rp{GM}{r}-\gamma(\bm v\cdot\bm v)
\right]\bm r+2(1+\gamma)(\bm r\cdot\bm v)\bm
v\right\},\lb{age}\eqf where $G$ is the Newtonian gravitational
constant, $c$ is the speed of light in vacuum, $\bm r$ and $\bm v$
are the position and velocity vectors, respectively, of the test
body, $\gamma$ and $\beta$ are the standard
Eddington-Robertson-Schiff PPN parameters ($\gamma=\beta=1$ in
GTR; in this case \rfr{age} reduces to the expression of (Mashhoon
et al. 2001)). Note that $r$ here is the standard isotropic radial
coordinate, not to be confused with the Schwarzschild radial
coordinate $r^{'}=r[1+GM/(2c^2 r)]^2$. In the usual orbital data
reductions it is just $r$ which is employed.

By considering it as a small perturbation to the Newtonian
monopole acceleration it is possible to work out its effect on the
orbital motion of  a test body with the standard perturbative
techniques. The secular rate of the argument of pericentre is
given by the well known formula \eqi\dot\omega_{\rm
GE}=\frac{3nGM}{c^2 a(1-e^2)}\nu_{\rm GE },\lb{ge}\eqf where
 \eqi\nu_{\rm GE }=\frac{2+2\gamma-\beta}{3},\lb{nuge}\eqf
$a$ and $e$ are the semimajor axis and the eccentricity,
respectively, of the test particle's orbit and $n=\sqrt{GM/a^3}$
is the (unperturbed) Keplerian mean motion. The orbital period of
an unperturbed, two-body Keplerian ellipse is $P=2\pi/n$. Note
that \rfr{ge} is an exact result valid to all order in $e$.

As we will show, also the mean anomaly $\mathcal{M}$ is affected
by a post-Newtonian gravitoelectric secular rate. The Gauss
perturbative equation for $\mathcal{M}$ is given by
\eqi\rp{d\mathcal{M}}{dt}=n-\rp{2}{na}A_R\left(\rp{r}{a}\right)-\sqrt{1-e^2}\left(\rp{d\omega}{dt}+\cos
i\rp{d\Omega}{dt} \right),\lb{manom}\eqf where $\Omega$ and $i$
are the longitude of the ascending node and the inclination of the
test particle's orbit to the equator of the central mass,
respectively, and $A_R$ is the radial component of the perturbing
acceleration. In order to obtain the post-Newtonian
gravitoelectric secular rate of the mean anomaly one must consider
the radial component of \rfr{age}, insert it in the second term of
the right-hand-side of \rfr{manom}, evaluate it on the unperturbed
Keplerian ellipse, characterized by $r=a(1-e^2)/(1+e\cos f)$ and
by the radial and along-track components of the velocity vector
which are $v_{R}=nae\sin f/\sqrt{1-e^2}$, $v_{T}=na(1+e\cos
f)/\sqrt{1-e^2}$, respectively ($f$  is the orbiter's true
anomaly),  multiplying it by
\eqi\rp{dt}{P}=\rp{(1-e^2)^{3/2}df}{2\pi(1+e\cos f )^2},\eqf  and
integrating over one orbital revolution, i.e. from 0 to $2\pi$. It
turns out that the post-Newtonian gravitoelectric secular rate of
the mean anomaly is, to order $\mathcal{O}(e^2)$
\eqi\dot\mathcal{M}_{\rm GE}\sim-\rp{nGM}{c^2
a}\left[\left(2+4\gamma+3\beta\right)\left(1+\rp{e^2}{2}\right)+\left(2+\gamma\right)e^2\right].\lb{manomi}\eqf
We can define \eqi\mu_{\rm GE}=\rp{2+4\gamma+3\beta}{9}.\eqf It
turns out that the second term of \rfr{manomi} induces for Mercury
an additional shift of 1.789 arcseconds/century ($''$ cy$^{-1}$ in
the following), while for the other planets it is of the order of
$10^{-2}$--$10^{-4}$ $''$ cy$^{-1}$. Thus, for them the secular
rate of the mean anomaly can be written as
\eqi\dot\mathcal{M}_{\rm GE}\sim-\rp{9nGM}{c^2
a\sqrt{1-e^2}}\mu_{\rm GE}.\eqf However, as we will see later, the
mean anomaly of Mercury will not be used in the combined residuals
strategy outlined in the following.
For Mercury, the post-Newtonian gravitoelectric effect induced by
the Sun on $\omega$ and $\mathcal{M}$, according to GTR, amounts
to 42.980 $''$ cy$^{-1}$ and -127.949 $''$ cy$^{-1}$,
respectively. The most accurate estimate of the gravitoelectric
perihelion advance seems to be that obtained for Mercury by E.M.
Standish in 2000 with the DE405 ephemerides (Standish 1998) and
reported in (Pireaux and Rozelot 2003). He averaged the Mercury's
perihelion evolution over two centuries by using the DE405
ephemerides with and without the post-Newtonian accelerations.
Standish included in the force models also the solar oblateness
with $J_{2\odot}=2\times 10^{-7}$, so that the residuals for
Mercury accounted for the post-Newtonian effects only; the
determined shift was $42.98\pm 0.0023$ $''$ cy$^{-1}$. The same
approach for the mean anomaly (Standish E M private communication)
has yielded to $-130.003\pm 0.0027$ $''$ cy$^{-1}$ (See also Table
\ref{rates} later). Note that the quoted uncertainty do not come
from direct observational errors. They depend on the fact that in
the force models used in the numerical propagation many
astrodynamical parameters occur (masses of planets, asteroids,
etc.); their numerical values come from multiparameter fits of
real data and, consequently, are affected by observational errors.
Such numerical tests say nothing about if GTR is correct or not;
they just give an idea of what would be the obtainable accuracy
set up by our knowledge of the Solar System arena if the Einstein
theory of gravitation would be true.

In regard to the possibility of constructing time series of
planetary mean anomalies, it must be noted that a certain effort
would be required. Indeed, so far this orbital elements has never
been utilized, so that the partials, e.g., should be computed
(Pitjeva, private communication 2004).
\subsection{The gravitomagnetic Lense-Thirring effect}
Another post-Newtonian secular precession which affects not only
$\omega$ but also $\Omega$ is the Lense-Thirring effect (Lense \&
Thirring 1918) induced by the proper angular momentum $J$ of the
central body \eqia\dot\omega_{\rm LT}&=&\frac{-6GJ\cos i}{c^2 a^3
(1-e^2)^{3/2}}\mu_{\rm LT},\lb{gm}\\
\dot\Omega_{\rm LT}&=&\frac{2GJ}{c^2 a^3 (1-e^2)^{3/2}}\mu_{\rm
LT},\lb{gm2}\eqfa where \eqi\mu_{\rm LT}=\frac{1+\gamma}{2}.\eqf
By assuming for the Sun $J_{\odot}=1.9\times 10^{48}$ g cm$^2$
s$^{-1}$ (Pijpers 2003), the gravitomagnetic effect on, e.g.,
Mercury's perihelion is of the order of $10^{-3}$ $''$ cy$^{-1}$
(De Sitter 1916). Such small value is at the edge of the present
sensitivity (0.002 $''$ cy$^{-1}$) in determining the Mercury
perihelion shift from the ephemerides (cfr. the results obtained
by Standish). Moreover, it should be considered that, even if
future improvements of the obtainable experimental sensitivity in
interplanetary ranging allowed to consider the possibility of
measuring the Lense-Thirring perihelion advance of Mercury, the
impact of the systematic errors due to the uncertainties in the
solar oblateness would severely limit the realistic accuracy
obtainable in such demanding measurement. Indeed, by assuming an
uncertainty of $\sigma_{J_{2\odot}}=0.4\times 10^{-7}$, the error
in the secular precession induced by the Sun's quadrupole moment
would amount to 245$\%$ of the Lense-Thrring shift of $\pi^{\rm
Merc}$ (see below for the definition of $\pi$). Another important
source of systematic error in the measurement of such a tiny
effect would be represented by the $N-$body classical secular
precessions which are of the order of $10^2-10^3$ $''$ cy$^{-1}$
(see on the WEB
\texttt{http://ssd.jpl.nasa.gov/elem$\_$planets.html}). Indeed,
their residual mismodelled part could severely bias the recovery
of the Lense-Thirring effect. For a recent approach to these
problems and possible strategies to overcome them, see (Iorio
2004). It must be noted that, if, on the other hand, we look at a
Lense--Thirring test as a way to measure the Sun's angular
momentum by assuming the validity of GTR, a measurement of the
solar gravitomagnetic field would have a significance only if the
obtainable accuracy was better than 10$\%$; indeed, among other
things, the present-day uncertainty in the Sun's angular momentum
$J_{\odot}$, which could be measured from the Lense-Thirring
precessions, is just of the order of 10$\%$ (Xu \& Ni 1997) in
various solar models (Patern\'{o} et al. 1996; Elsworth et al.
1995) or even less in the framework of asteroseismogyrometry
(Pijpers 2003).

At present, the only performed attempts to explicitly extract the
Lense-Thirring signature from the data of orbiting masses in the
Solar System are due to Ciufolini and coworkers who analyzed the
laser-ranged data of the orbits of the existing LAGEOS and LAGEOS
II Earth artificial satellites (Ciufolini et al. 1998). A
20$-30\%$ precision level in measuring the terrestrial
gravitomagnetic field is claimed, but other scientists judge these
evaluations too optimistic and propose different error budget
(Ries et al. 2003). In April 2004 the GP-B spacecraft has been
launched. It will carry out a very complex and challenging mission
which should be able to measure a gravitomagnetic precession of
the spins of four superconducting gyroscopes (Schiff 1960) carried
on board at a claimed accuracy of 1\% or better (Everitt et al.
2001).
\section{The solar oblateness}
The solar quadrupole mass moment $J_{2\odot}$ is an important
astrophysical parameter whose precise knowledge could yield many
information about the inner structure and dynamics of our star. A
reliable evaluation of $J_{2\odot}$ still faces some controversy:
on one side, the theoretical values strongly depend on the solar
model used, whereas accurate measurements are very difficult to
obtain from observations. For all this matter see the recent
review (Pireaux \& Rozelot 2003) and (Rozelot et al. 2004). From
an observational point of view, $J_{2\odot}$ is not directly
accessible. In this context a dynamical determination of
$J_{2\odot}$, analyzing, e.g., the orbits of the inner planets of
the Solar System, is interesting because it might be compared with
those derived from solar model dependent values of the oblateness.
However, it is not simple to reach this goal because of the
interplay between the effects of the solar quadrupole moment with
those induced by the post-Newtonian gravitoelectromagnetic forces.
Instead of the trajectory of planets, it would be possible to
infer $J_{2\odot}$ from accurate tracking of some drag-free
spacecraft orbiting within a few radii of the solar center. This
will be the approach followed by, e.g. the BepiColombo mission
(see Section \ref{newmiss}). Alternatively, the Sun's quadrupole
mass moment can be inferred from in-orbit measurement of solar
properties, like the SOHO-$MDI$ space-based observations
(Armstrong \& Kuhn 1999), or from Earth-based observations like
those realized, e.g., with the scanning heliometer of the Pic du
Midi Observatory (Rozelot et al. 2004).
\subsection{The classical precessions induced by the solar oblateness}
For a given planet of the Solar System orbiting the Sun, apart
from the classical effects induced by the precession of the
equinoxes and by the other planets and major asteroids which are
routinely accounted for in the ephemerides computations (Pitjeva
2001a; 2001b), the oblateness of Sun induces also secular
precessions on $\Omega,\ \pi=\omega+\Omega\cos i$ and
$\mathcal{M}$ given by \eqia
\dot\Omega_{J_{2\odot}}&=&-\frac{3}{2}\frac{nJ_{2\odot}}{(1-e^2)^2}\left(\frac{R_{\odot}}{a}\right)^2\cos i,\\
\dot\pi_{J_{2\odot}}&=&-\frac{3}{2}\frac{nJ_{2\odot}}{(1-e^2)^2}\left(\frac{R_{\odot}}{a}\right)^2\left(\frac{3}{2}\sin^2
i-1 \right),\lb{obla}\\\lb{oblaanom}
\dot\mathcal{M}_{J_{2\odot}}&=&\frac{3}{4}\frac{nJ_{2\odot}
}{(1-e^2)^{3/2}}\left(\frac{R_{\odot}}{a}\right)^2 (3\cos^2 i-1),
\eqfa For Mercury\footnote{As pointed out in (Milani et al.,
2002), the angle $i$ refers to the inclination between the
planet's orbital plane and the fixed reference plane of the
celestial reference frame; it is not the angle $\epsilon$ between
the planet's orbital plane and the ecliptic. It turns out that
$i\sim\epsilon/2$. For Mercury $\epsilon=7.00487^\circ$. } the
Newtonian precessions due to Sun oblateness, with
$J_{2\odot}=2\times 10^{-7}$, are of the order of $10^{-2}$ $''$
cy$^{-1}$ for $\Omega$, $\pi$ and $\mathcal{M}$.
\section{The interplay between the solar oblateness and the
post-Newtonian precessions}
In Table \ref{tab1} the relevant parameters of the classical and
post-Newtonian secular precessions of the nodes, the perihelia and
the mean anomalies of Mercury, Venus and Mars are reported.
\begin{table}
\caption[]{Post-Newtonian precessions and coefficients of
Newtonian precessions of the node, the perihelion and the mean
anomaly for Mercury, Venus and Mars in $''$ cy$^{-1}$. The values
$\dot\pi_{\rm GE}$, and $\dot\mathcal{M}_{\rm GE}$ are calculated
with GTR. The coefficients $\dot\Omega_{.2}$, $\dot\pi_{.2}$ and
$\dot\mathcal{M}_{.2}$ are
$\partial(\dot\Omega_{J_{2\odot}})/\partial(J_{2\odot})$,
$\partial(\dot\pi_{J_{2\odot}})/\partial(J_{2\odot})$ and
$\partial(\dot\mathcal{M}_{J_{2\odot}})/\partial(J_{2\odot})$,
respectively. In order to have the precessions they must be
multiplied by $J_{2\odot}$. Note that the result for the mean
anomaly of Mercury accounts for the correction of order
$\mathcal{O}(e^2)$ which, instead, can be neglected for the other
planets. } \label{tab1}
$$
\begin{array}{p{0.5\linewidth}llllll}
            \hline
            \noalign{\smallskip}
Planet  & \dot\Omega_{\rm GE} & \dot\pi_{\rm GE} &
\dot\mathcal{M}_{\rm GE} & \dot\Omega_{.2}
& \dot\pi_{.2} & \dot\mathcal{M}_{.2} \\
\noalign{\smallskip}
            \hline
            \noalign{\smallskip}
Mercury & 0 & 42.981 & -127.949 & -126878.626 & 126404.437 & 123703.132 \\
Venus & 0 & 8.624 & -25.874 & -13068.273 & 13056.803 & 13056.504\\
Mars & 0 & 1.351 & -4.035 & -980.609 & 980.353 & 976.067\\
\noalign{\smallskip}
            \hline
         \end{array}
     $$
\end{table}
In all the relativistic tests performed up to now by analyzing the
perihelia advances only of the inner planets of the Solar System
with the radar ranging technique (Shapiro et al. 1972; 1976;
Shapiro 1990) it has been impossible to disentangle the genuine
post-Newtonian gravitoelectric contribution of \rfr{ge} from the
Newtonian precession of \rfr{obla}. Indeed, the observational
residuals of $\dot\pi$ for a single planet, built up by suitably
switching off the post-Newtonian $\mathcal{O}(c^{-2})$ terms and
the oblateness of Sun in the force models of the equations of
motion in the orbital processors softwares, account entirely for
the post-Newtonian and the solar oblateness\footnote{Also if the
solar oblateness is included in the force models, the related
uncertainty induces a corresponding systematic error in the
recovered post-Newtonian effect. Fortunately, it is small; by
assuming $\sigma_{J_{2\odot}}=0.4\times 10^{-7}$, it amounts to
0.01$\%$.} effects. This is a unsatisfactory situation, both if we
are interested in testing post-Newtonian gravity and if we want to
obtain a dynamical, model-independent measurement of $J_{2\odot}$.
Indeed, it is, of course, impossible to constraint both the
effects if only one perihelion rate is examined one at a time: in
recovering one of the two effects we are forced to consider the
other one as if it was known. Since the post-Newtonian
gravitoelectric effect is three orders of magnitude larger than
that induced by solar oblateness, a determination of the latter by
assuming the validity of GTR would be affected by a non negligible
systematic error induced by the precision to which the
post-Newtonian pericentre advance is known from other (more or
less indirect and more or less biased by other aliasing effects)
tests (Lunar Laser Ranging, binary pulsars periastron
advance\footnote{Note that the binary pulsars periastron
measurement should not be considered as a test of relativistic
gravity because the masses of the binary system are not known
(Stairs et al. 1998); they can be obtained by assuming the
validity of GTR. The Lunar Laser Ranging (LLR) measurements do not
allow to single out uniquely the gravitoelctric pericentre advance
from the other post-Newtonian features of motion of the Earth-Moon
system (Nordtvedt 2001). Indeed, LLR tests the post-Newtonian
equations of motion as a whole. Recently, it has been proposed to
measure the relativistic gravitoelectric perigee advance of the
terrestrial LAGEOS II satellite (Iorio et al. 2002; Lucchesi
2003), but, up to now, the test has not yet been performed. }).
The inverse situation is more favorable: indeed, if we are
interested in the post--Newtonian gravitoelectric effect the
relative systematic error induced on its measurement by the
precession due to the solar oblateness amounts to $5\times
10^{-4}$ even by assuming for the latter effect a 100$\%$
uncertainty\footnote{Note that the impossibility of disentangle
the gravitoelectric and Lense-Thirring effects would not seriously
affect the recovery of the gravitoelctric precession: indeed, the
bias induced by the gravitomagnetic effect on the gravitoelectric
shift amounts to 0.004$\%$ only for Mercury.}.
\subsection{The present-day approach to test post-Newtonian gravity}
At this point it may be interesting to clarify what is the current
approach in testing post-Newtonian gravity from planetary data
analysis followed by, e.g., the Jet Propulsion Laboratory (JPL).
In the interplay between the real data and the equations of
motions, which include also the post-Newtonian accelerations
expressed in terms of the various PPN parameters, a set of
astrodynamical parameters, among which there are also $\gamma$ and
$\beta$, are simultaneously and straightforwardly fitted and
adjusted and a correlation matrix is also released. This means
that the post-Newtonian equations of motion are globally tested as
a whole in terms of, among other parameters, $\gamma$ and $\beta$;
no attention is paid to this or that particular feature of the
post-Newtonian accelerations. The point is that the standard PPN
formalism refers to the alternative theories of gravitation which
are metric, i.e. based on a symmetric spacetime metric. But it is
not proven that an alternative theory of gravitation must
necessarily be a metric one. Moreover, even in the framework of
the metric alternative theories, the PPN formalism based on 10
parameters is not sufficient to describe every conceivable metric
theory of gravitation at the post-Newtonian order; it only
describes those theories with a particularly simple post-Newtonian
limit. One would, in principle, need an infinite set of new
parameters to add to the standard ten parameter PPN formalism in
order to describe the post-Newtonian approximation of any a priori
conceivable metric theory of gravity (Ciufolini 1991; Ciufolini \&
Wheeler 1995).
\subsection{The possibilities opened by the future missions}\lb{newmiss}
Concerning the possibility of disentangle the effects of the solar
oblateness from those of the post-Newtonian gravitoelectric force,
it is stated that the future space mission
BepiColombo\footnote{See on the WEB
\texttt{http://astro.estec.esa.nl/BepiColombo/}. Present ESA plans
are for a launch in 2010-2012.} of the European Space Agency (ESA)
will provide us, among other things, with a dynamical,
model-independent and relativity-independent measurement of
$J_{2\odot}$ by measuring with high precision the nodal motion of
Mercury (Milani et al. 2002; Pireaux \& Rozelot 2003) which is not
affected by the post-Newtonian gravitoelectric force. The claimed
accuracy would amount to $\sigma_{J_{2\odot}}=2\times 10^{-9}$
(Milani et al. 2002). However, such evaluation refers to the
formal, statistical obtainable uncertainty only. Indeed, the
residuals of the Mercury's node would account, to a certain level
of accuracy, for the Lense-Thirring precession as well. By
considering such effect as totally unmodelled in the force models,
its impact on the measurement of $J_{2\odot}$ would induce a
$8\times 10^{-9}$ systematic error. The formal, statistical
accuracy for $\gamma$ and $\beta$ is evaluated to be of the order
of $2\times 10^{-6}$ (Milani et al. 2002). Also the ESA
astrometric mission GAIA\footnote{See on the WEB
\texttt{http://astro.estec.esa.nl/GAIA/}. Present ESA plans are
for a launch in mid-2010.} should measure, among other things, the
solar quadrupole mass moment by analyzing the longitudes of the
ascending nodes of many minor bodies of the Solar System. The
obtainable accuracy for $\gamma$ is of the order of
$10^{-5}$-10$^{-7}$ (Vecchiato et al. 2003). The ASTROD mission
should be able to measure $J_{2\odot}$ with a claimed accuracy of
the order of $10^{-8}$ or, perhaps, $10^{-9}-10^{-10}$ (Ni et al.
2004). The claimed obtainable accuracy for the PPN parameters is
$4.6\times 10^{-7}$ for $\gamma$ and $4\times 10^{-7}$ for
$\beta$. Further improvements may push these limits down to
$10^{-8}-10^{-9}$. The recently proposed LATOR mission should be
able to measure, among other things, $\gamma$ to a 10$^{-8}$
accuracy level and $J_{2\odot}$ to a $10^{-8}$ level (Turyshev et
al. 2004).
The solar
orbit relativity test SORT (Melliti et al. 2002), which would
combine a time-delay experiment with a light deflection test,
should allow to reach a $10^{-6}$ accuracy in measuring the PPN
parameters
\section{The experimental accuracy in planetary radar ranging}
Concerning the present-day accuracy of the planetary radar
ranging, the radar itself is accurate well below the 100 m level
(Standish 2002). The problem, however, comes from the fact that
the surfaces of the planets have large topographical variations.
They are modeled in different ways. For Mercury, spherical
harmonics and some closure analysis (comparing values when two
different measurements reflect off from the same spot on the
surface) have been done (Anderson et al. 1996) in DE405. For
Venus, a topographical model, which comes from (Pettengill et al.
1980), has been used. For Mars, closure points can be used.
Closure points are pairs of days during which the observed points
on the surface of Mars are nearly identical with respect to their
longitudes and latitudes on Mars. Since the same topographical
features are observed during each of the two days, the uncertainty
introduced by the topography may be eliminated by subtracting the
residuals of one day from the corresponding ones of the other day.
The remaining difference is then due to only the ephemeris drift
between the two days. The closure points for Mars have a priori
uncertainties of about 100 m or less when the points are within
0.2 degrees of each other on the martian surface. Of course, for
Mars, there is also the spacecraft ranging - far more accurate
than the radar: Viking Landers (1976-82), 10 m; Mars Global
Surveyor and Odyssey (1999-2003), 2-3 m. However, correction for
Mars topography is possible not only by using closure points (in
this method no all observations may be used), but with help of
modern hypsometric maps and by the representations of the global
topography with an expansion of spherical functions (Pitjeva
2001b). It must now be noted that our knowledge of the orbital
motion of Mercury should improve thanks to the future hermean
missions Messenger (see on the WEB http://messenger.jhuapl.edu/
and http://discovery.nasa.gov/messenger.html), which has been
launched in the summer 2004 and whose encounter with Mercury is
scheduled for 2011, and, especially\footnote{While the spacecraft
trajectory will be determined from the range-rate data, the
planet's orbit will be retrieved from the range data (Milani et
al. 2002). In particular, the determination of the planetary
centre of mass is important to this goal which can be better
reached by a not too elliptical spacecraft's orbit. The relatively
moderate ellipticity of the planned 400$\times$ 1500 km polar
orbit of BepiColombo, contrary to the much more elliptical path of
Messenger, is, then, well adequate.}, BepiColombo. A complete
error analysis for the range and range-rate measurements can be
found in (Iess \& Boscagli 2001). According to them, a full 5-way
link to the main orbiter will be adopted. A multi-channel
combination of the data will allow to remove most of the
measurement errors introduced, in a single channel, by the plasma.
As a result, a two orders of magnitude improvement in the
Earth-Mercury range should be possible, also avoiding the problems
related to the surface topography. According to a more
conservative evaluation by E.M Standish (Standish, private
communication 2004), improvements in the Mercury's orbital
parameters might amount to one order of magnitude, i.e. the tens
of meters level.
\begin{table}
\caption[]{Present-day accuracy in determining the node,
perihelion and mean anomaly secular rates of Mercury, Venus and
Mars according to DE405 (Standish 1998) and EPM2000 (Pitjeva
2001a; 2001b) ephemerides. The figures, in $''$ cy$^{-1}$,
represent the formal, statistical errors. Realistic errors should
be 10 times larger, at least. While the results by Standish come
from the mathematical propagation  of the nodes, the perihelia and
the mean anomalies (Standish, E. M. private communication)
evolution with and without post--Newtonian terms (with
$\gamma=\beta=1$) and their average over a time span of two
centuries, the results by Pitjeva are based on real data. The
figure for Mercury has been obtained in 2001, while the other ones
have been determined subsequently (Pitjeva, personal communication
2004). } \label{rates}
$$
         \begin{array}{p{0.5\linewidth}llll}
            \hline
            \noalign{\smallskip}
Planet  & \sigma_{\dot\Omega_{\rm calc}} (DE405) &
\sigma_{\dot\pi_{\rm calc}} (DE405) & \sigma_{\dot\mathcal{M}_{\rm
calc}}
(DE405) & \sigma_{\dot\pi_{\rm obs}} (EPM2000) \\
\noalign{\smallskip}
            \hline
            \noalign{\smallskip}
Mercury & 0.000182 & 0.0023 & 0.0027 & 0.0086\\
Venus & 0.000006 & 0.0414 & 0.0414 & 0.1037\\
Mars & 0.000001 & 0.0014 & 0.0014 & 0.0001\\
\noalign{\smallskip}
            \hline
         \end{array}
     $$
\end{table}
\section{The proposed approach for disentangle the solar oblateness and the post-Newtonian effects}
Let us write down the following equations\footnote{Note that in
the right-hand-sides of \rfrs{uno}{due} also the mismodelled parts
of the classical $N-$body precessions should have been included.
Since they are small with respect to the gravitoelectric effects
of interest, as we will show later, we can neglect them in the
calculations for setting up our combiantions. This, of course,
does not mean that the experimental residuals do not account also
for them.}

\begin{equation}
\left\{
\begin{array}{lll}\lb{geidue}
\delta\dot\Omega^{\rm Merc}_{\rm obs}&=&\dot\Omega^{\rm
Merc}_{.2}J_{2\odot}+\dot\Omega^{\rm Merc}_{\rm
N-body}+\dot\Omega^{\rm Merc}_{\rm LT}\mu_{\rm
LT},\\\\
\delta\dot\Omega^{\rm Venus}_{\rm obs}&=&\dot\Omega^{\rm
Venus}_{.2}J_{2\odot}+\dot\Omega^{\rm Venus}_{\rm
N-body}+\dot\Omega^{\rm Venus}_{\rm LT}\mu_{\rm
LT},\\\\
\delta\dot\Omega^{\rm Mars}_{\rm obs}&=&\dot\Omega^{\rm
Mars}_{.2}J_{2\odot}+\dot\Omega^{\rm Mars}_{\rm
N-body}+\dot\Omega^{\rm Mars}_{\rm LT}\mu_{\rm
LT},\\\\
\end{array}
\right.
\end{equation}

\begin{equation}
\left\{
\begin{array}{lll}\lb{uno}
\delta\dot\mathcal{M}^{\rm Mars}_{\rm obs}&=&\dot\mathcal{M}^{\rm
Mars}_{.2}J_{2\odot}+\dot\mathcal{M}^{\rm Mars}_{\rm GE}\mu_{\rm
GE},\\\\
\delta\dot\mathcal{M}^{\rm Venus}_{\rm obs}&=&\dot\mathcal{M}^{\rm
Venus}_{.2}J_{2\odot}+\dot\mathcal{M}^{\rm Venus}_{\rm GE}\mu_{\rm
GE},\\
\end{array}
\right.
\end{equation}

\begin{equation}
\left\{
\begin{array}{lll}\lb{due}
\delta\dot\pi^{\rm Mars}_{\rm obs}&=&\dot\pi^{\rm
Mars}_{.2}J_{2\odot}+\dot\pi^{\rm Mars}_{\rm GE}\nu_{\rm
GE},\\\\
\delta\dot\pi^{\rm Merc}_{\rm obs}&=&\dot\pi^{\rm
Merc}_{.2}J_{2\odot}+\dot\pi^{\rm Merc}_{\rm GE}\nu_{\rm
GE},\\
\end{array}
\right.
\end{equation}
where $\delta\dot\Omega^{\rm Planet}_{\rm obs}$,
$\delta\dot\pi^{\rm Planet}_{\rm obs}$ and
$\delta\dot\mathcal{M}^{\rm Planet}_{\rm obs}$ are the
observational residuals\footnote{Here we speak about residuals of
Keplerian orbital elements in a, strictly speaking, improper
sense. The Keplerian orbital elements are not directly observable:
they can only be computed. The basic observable quantities are
ranges, range-rates and angles. Here we mean the differences
between the time series of the node got from a given observed
orbital arc and the time series of the node got from a propagated
orbital arc with a given force, which we are interested in,
switched off in the force models. The two time series share the
same initial conditions.} of the rates of the nodes, the
longitudes of the perihelia and the mean anomalies of Mercury,
Venus and Mars. It is intended that all kind of data (optical and
radio) would be used. The residuals should be built up by
purposely switching off the solar quadrupole moment and the
post-Newtonian gravitoelectric accelerations in the force models
(or leaving in them some default values to be subsequently
adjusted according to the present strategy) of the orbital
processors. Then, the so obtained observational residuals would
entirely (or partly, if some default values are left in the force
models) adsorb just the investigated secular effects and other
post--Newtonian short--periodic features, i.e. not averaged over
one orbital revolution of the planet under consideration. In
respect to the latter point, it should be noted that the residuals
of the mean anomalies would account, e.g., also for the indirect
effects on the mean motions $n$ through the perturbations in the
semimajor axes $a$ \eqi\Delta
n=-\rp{3}{2}\sqrt{\rp{GM}{a^5}}\Delta a\lb{deltan}.\eqf  There are
no secular perturbations induced on $a$ by the other planets. If
the classical short--periodic effects on $a$ would be of
relatively no importance because they would be included in the
force models at the best of their accuracy, this is not the case
for the post-Newtonian ones. The gravitoelectric field induces no
secular variations on $a$, as the classical planetary
perturbations. The short--term shift on $a$ can be calculated from
\rfr{age} and the Gauss equation for the perturbed rate of
semimajor axis \eqi \frac{da}{dt}=\rp{2}{n\sqrt{1-e^2}}\left[A_R
e\sin f+A_T(1+e\cos f )\right], \eqf where $A_T$ is the
along-track component of the perturbing acceleration. It amounts
to \eqi \Delta a_{\rm GE}=\rp{GMe}{c^2 (1-e^2)^2}[14(\cos f_0-\cos
f)+10e(\cos^2 f-\cos^2 f_0 )]+\mathcal{O}(e^3).\eqf For Venus
their nominal amplitudes are of the order of 100 m; for Mars they
amount to 2 km; \rfr{deltan} would yield periodic variations whose
nominal amplitudes would be of the order of 0.2-1 $''$ cy$^{-1}$
for Venus and Mars, respectively. However, such harmonic
signatures could be fitted and removed from the data over
sufficiently long time spans.
Indeed, since we are interested in the gravitoelectric secular
trends on $\pi$ and $\mathcal{M}$ it should be possible, in
principle, to construct the residuals by using orbital arcs longer
than the sidereal revolution periods of the planets to be used.
Then, all the high--frequency perturbations would not affect their
time series which should, instead, be characterized by the secular
parts of those Newtonian and post--Newtonian features present in
the real data but (partly) absent in the force models of the
equations of motion in the orbital processors. However, in regard
to the possibility of constructing accurate time series of
observational residuals many years long the following observations
must be kept in mind (Standish 2002). The planetary motions are
perturbed by the presence of many asteroids whose masses are quite
poorly known. Furthermore, it is not possible to solve for the
asteroid masses, other than for the biggest few, because there are
too many of them for the data to support such an effort. As a
result, the ephemerides of the inner planets, especially Mars,
will deteriorate over time; the ephemerides have uncertainties at
the 1-2 km level over the span of the observations and growing at
the rate of a few km/decade outside that span. On the other hand,
it must also be noted that the sidereal orbital periods of Venus
and Mars amount to 224.701 and 686.980 days, respectively.

We can consider \rfrs{geidue}{due} as three systems of algebraic
linear equations in the three unknowns\footnote{This approach is
analogous to that employed in the LAGEOS-LAGEOS II Lense-Thirring
experiment in the gravitational field of Earth (Ciufolini 1996;
Ciufolini et al. 1998)} $J_{2\odot}, \mu_{\rm GE}$ and $\nu_{\rm
GE}$. Their solutions can be written as
\begin{equation}
\left\{
\begin{array}{lll}\lb{tre}
\delta\dot\Omega^{\rm Merc}_{\rm obs}+c_1\delta\dot\Omega^{\rm
Venus}_{\rm obs}+c_2\delta\dot\Omega^{\rm Mars}_{\rm obs}&=&
J_{2\odot}\left(\dot\Omega^{\rm
Merc}_{.2}+c_1\dot\Omega^{\rm Venus}_{.2}+c_2\dot\Omega_{.2}^{\rm Mars}\right),\\\\
\delta\dot\mathcal{M}^{\rm Mars}_{\rm
obs}+c^{'}_1\delta\dot\mathcal{M}^{\rm Venus}_{\rm obs}&=&\mu_{\rm
GE}\left(\dot\mathcal{M}^{\rm
Mars}_{\rm GE}+c^{'}_1\dot\mathcal{M}^{\rm Venus}_{\rm GE}\right),\\\\
\delta\dot\pi^{\rm Mars}_{\rm obs}+c^{''}_1\delta\dot\pi^{\rm
Merc}_{\rm obs}&=& \nu_{\rm GE}\left(\dot\pi^{\rm
Mars}_{\rm GE}+c^{''}_1\dot\pi^{\rm Merc}_{\rm GE}\right),\\
\end{array}
\right.
\end{equation}
where
\begin{equation}
\left\{
\begin{array}{lll}\lb{quattro}
c_1&=&\frac{\dot\Omega^{\rm Mars}_{\rm LT}\dot\Omega^{\rm
Merc}_{\rm N-body } -\dot\Omega^{\rm Merc}_{\rm LT}\dot\Omega^{\rm
Mars}_{\rm N-body }}{\dot\Omega^{\rm Venus}_{\rm
LT}\dot\Omega^{\rm Mars}_{\rm N-body }-\dot\Omega^{\rm Mars}_{\rm
LT}\dot\Omega^{\rm Venus}_{\rm N-body}}=-7.73247,\

c_2=\frac{\dot\Omega^{\rm Merc}_{\rm LT}\dot\Omega^{\rm
Venus}_{\rm N-body} -\dot\Omega^{\rm Venus}_{\rm
LT}\dot\Omega^{\rm Merc}_{\rm N-body }}{\dot\Omega^{\rm
Venus}_{\rm LT}\dot\Omega^{\rm Mars}_{\rm N-body }-\dot\Omega^{\rm
Mars}_{\rm LT}\dot\Omega^{\rm Venus}_{\rm
N-body}}=7.11840\\\\
c^{'}_1&=&-{\dot\mathcal{M}^{\rm Mars}_{.2}}/{\dot\mathcal{M}^{\rm Venus}_{.2}}=-0.07475,\\\\
c^{''}_1&=&-{\dot\pi^{\rm Mars}_{.2}}/{\dot\pi^{\rm Merc}_{.2}}=-0.00775,\\
\end{array}
\right.
\end{equation}
and
\begin{equation}
\left\{
\begin{array}{lll}\lb{cinque}
\dot\Omega^{\rm Merc}_{.2}+c_1\dot\Omega^{\rm
Venus}_{.2}+c_2\dot\Omega^{\rm Mars}_{.2}&=&-32808.8816\ ''\ {\rm cy}^{-1},\\\\
\dot\mathcal{M}^{\rm Mars}_{\rm
GE}+c^{'}_1\dot\mathcal{M}^{\rm Venus}_{\rm GE}&=& -2.1007\ ''\ {\rm
cy}^{-1},\\\\
\dot\pi^{\rm Mars}_{\rm GE}+c^{''}_1\dot\pi^{\rm Merc}_{\rm
GE}&=&1.0176\ ''\ {\rm cy}^{-1}\\
\end{array}
\right.
\end{equation}
The first equation of \rfr{tre} comes from \rfr{geidue} solved for
$J_{2\odot}$; it allows to obtain $J_{2\odot}$ independently of
the post-Newtonian Lense-Thirring and classical $N-$ body secular
precessions which would represent the major sources of systematic
errors. The second equation of \rfr{tre} comes from \rfr{uno}
solved for $\mu_{\rm GE}$; it cancels out the secular precessions
due to $J_{2\odot}$. The third equation in \rfr{tre} comes from
\rfr{due} solved for $\nu_{\rm GE }$; it cancels out the secular
precessions due to $J_{2\odot}$. In regard to the impact of the
residual mismodelled classical $N-$ body precessions on the second
and the third combinations of \rfr{tre}, they should not induce a
systematic error larger than the observational one (see below)
because the expected values of $\mu_{\rm GE}$ and $\nu_{\rm GE}$
are of the order of unity\footnote{Let us quantitatively discuss
this point. By using the results for the observed centennial rates
$\dot\Omega$ and $\dot\varpi$ released at
\texttt{http://ssd.jpl.nasa.gov/elem$\_$planets.html} in, say, the
left-hand-side of the third combination of \rfr{tre} it is
possible to obtain for it a nominal $N-$body shift of 539.6036
$''$ cy$^{-1}$. The uncertainty in the $N-$body precessions lies
mainly in the $Gm$ of the perturbing planets, among which Jupiter
plays the major role. Now, the relative uncertainty in Jupiter's
$Gm$ is of the order of 10$^{-9}$ m$^3$ s$^{-2}$ (Jacobson 2003);
then, a reasonable estimate of the order of magnitude of the
mismodelled part of the $N-$body shift should be $1\times 10^{-6}$
$''$ cy$^{-1}$. This figure must be divided by 1.0176 $''$
cy$^{-1}$ yielding a relative error in $\nu_{\rm GE}$ of the order
of $10^{-6}$. Cfr. with \rfr{sette}.}, contrary to $J_{2\odot}$
which should be of the order of $10^{-7}$.  The adimensional
parameters $J_{2\odot},\ \mu_{\rm GE}$ and $\nu_{\rm GE}$ are
estimated by fitting the time series of the left-hand-sides of
\rfr{tre} with straight lines, measuring their slopes, in $''$
cy$^{-1}$, and, then, by dividing them by the the quantities of
\rfr{cinque} which have the dimensions of $''$ cy$^{-1}$. Note
that the solar quadrupole mass moment would not be affected by the
indirect effects on $n$ because only the nodes would be used in
its determination. Finally, from the so obtained values of
$\mu_{\rm GE}$ and $\nu_{\rm GE}$, which are 1 in GTR and 0 in
Newtonian mechanics, it is possible to measure $\gamma$ and
$\beta$ independently of the solar oblateness and also of each
other as
\begin{equation}
\left\{
\begin{array}{lll}\lb{sei}
\gamma&=&\frac{9}{10}(\mu_{\rm GE}+\nu_{\rm GE})-\frac{4}{5},\\\\
\beta&=&\rp{9}{5}\mu_{\rm GE}-\rp{6}{5}\nu_{\rm GE}+\rp{2}{5}.\\
\end{array}
\right.
\end{equation}
According to the results of Table \ref{rates} it is possible to
yield an estimate of the (formal) uncertainty in $J_{2\odot},\
\mu_{\rm GE}$ and $\nu_{\rm GE}$ as
\begin{equation}
\left\{
\begin{array}{lll}\lb{sette}
\sigma_{J_{2\odot}}&\sim& 4.5\times 10^{-9},\\\\
\sigma_{\mu_{\rm GE}}&\sim& 6\times 10^{-5},\\\\
\sigma_{\nu_{\rm GE}}&\sim& 3\times 10^{-5}.\\
\end{array}
\right.
\end{equation}
The evaluation for $\sigma_{\mu_{\rm GE}}$ accounts also for the
fact that \rfr{deltan} and the values of $\sigma_a$ of Table VI of
(Pitjeva 2001a) yield a (formal) uncertainty of $3\times 10^{-5}$.
Note that the accuracies in $\nu_{\rm GE}$ and $J_{2\odot}$ would
be further improved from the better knowledge of Mercury's orbit
which should be obtained from the BepiColombo mission. The
uncertainties in $\gamma$ and $\beta$ would amount to
\begin{equation}
\left\{
\begin{array}{lll}\lb{otto}
\sigma_{\gamma}&\sim& 6\times 10^{-5},\\\\
\sigma_{\beta}&\sim& 1\times 10^{-4}.\\
\end{array}
\right.
\end{equation}
However, discretion is advised in evaluating the reliability of
these results because they refer to the formal, standard
statistical errors; realistic errors may be also one order of
magnitude larger.

For the most recent determinations of $\gamma$ and $\beta$ we have
that, according to the frequency shift of radio photons to and
from the Cassini spacecraft, $\sigma_{\gamma}=2.3\times 10^{-5}$
(Bertotti et al. 2003). This result, combined with
$\sigma_{\eta}=4.5\times 10^{-4}$ for the Strong Equivalence
Principle (SEP) violating parameter $\eta=4\beta-\gamma -3$
(Nordtvedt 1968a; 1968b) from the most recent analysis of LLR
data, yields $\sigma_{\beta}=1\times 10^{-4}$ (Turyshev et al.
2003). Conservative results from previous LLR analyses yield
realistic uncertainties\footnote{The SEP parameter uncertainty in
(M\"{u}ller et al. 1998) is $\sigma_{\eta}=9\times 10^{-4}$.}
(M\"{u}ller et al. 1998) $\sigma_{\gamma}=5\times 10^{-3}$ and
$\sigma_{\beta}=4\times 10^{-3}$. Recent radar ranging
measurements to planets yield (Pitjeva 2003)
$\sigma_{\gamma}=1\times 10^{-4}$ and $\sigma_{\beta}=1\times
10^{-4}$. However, as pointed out in (Pitjeva 2003),  these
uncertainties are formal standard deviations; realistic error
bounds may be an order of magnitude larger.
\section{Conclusions}
In this paper we have explicitly worked out the post--Newtonian
gravitoelectric secular rate of the mean anomaly of a test
particle freely orbiting a spherically symmetric central object.
Moreover, we have outlined a possible strategy for determining
simultaneously and independently of each other, the solar
quadrupole mass moment $J_{2\odot}$ and two parameters $\nu_{\rm
GE}$ and $\mu_{\rm GE}$ which account for the post--Newtonian
gravitoelectric secular shifts of the perihelion and the mean
anomaly, respectively, of planets. They are 0 in Newtonian
mechanics and 1 in the General Theory of Relativity. They could be
expressed in terms of the standard PPN $\gamma$ and $\beta$
parameters which could, then, be determined independently of each
other as well. The usual approach employed in the ephemerides data
reductions tests post--Newtonian gravity theories as a whole
straightforwardly in terms of the PPN parameters involving the
simultaneous fit of many more or less correlated astrodynamical
parameters among which there are also $\gamma$, $\beta$ and
$J_{2\odot}$. In this case, instead, we propose to analyze the
time series of three suitably linear combinations of the residuals
of $\dot\Omega$, $\dot\pi$ and $\dot\mathcal{M}$ of Mercury, Venus
and Mars built up in order to single out just certain selected
Newtonian and post--Newtonian orbital effects. This can be done by
setting purposely equal to zero (or to some default values to be
subsequently adjusted with the proposed strategy) the orbital
effects of interest in the force models of the equations of
motion. By suitably choosing the length of the orbital arcs it
would be possible to account for the secular terms only. The
coefficients of the combinations would make each of such
combinations sensitive just to one orbital effect at a time,
independently of the other ones. By fitting the experimental
residual signals with straight lines, measuring their slopes in
$''$ cy$^{-1}$ and suitably normalizing them would yield the
values of $J_{2\odot},\ \mu_{\rm GE}$ and $\nu_{\rm GE}$. The
obtainable formal accuracy would be of the order of
$10^{-4}-10^{-5}$ for $\gamma$ and $\beta$ and, more
interestingly, 10$^{-9}$ for $J_{2\odot}$. These estimates should
be improved by the future more accurate tracking data of Mercury's
orbit from the BepiColombo mission.
\begin{acknowledgements}
L. Iorio is grateful to E. M. Standish (JPL) and E. V. Pitjeva
(Institute of Applied Astronomy of Russian Academy of Sciences),
for their help and useful discussions and clarifications, and to
W.-T. Ni (Purple Mountain Observatory) for the updated reference
on ASTROD. Thanks also to S.Turyshev (JPL) for his important
clarifications on the LATOR mission.
\end{acknowledgements}

\end{document}